\documentclass[3p,times]{elsarticle}

\usepackage{ecrc}

\newcommand{\ket}[1]{|#1\rangle}

\usepackage{xspace}
\usepackage{color}
\usepackage[table,x11names]{xcolor}
{}
{}
{}

\usepackage{url}
\usepackage{multirow}
\usepackage{enumerate}
\usepackage{graphicx}
\usepackage{braket}
\usepackage{todonotes}
\usepackage{amsmath}
\usepackage{hyperref}
\usepackage{amsmath} 
\usepackage{comment}
\definecolor{LightCyan}{rgb}{0.88,1,1}
\volume{00}

\firstpage{1}

\journalname{Procedia Computer Science}

\runauth{}

\jid{procs}

\jnltitlelogo{Procedia Computer Science}

\CopyrightLine{2011}{Published by Elsevier Ltd.}

\usepackage{amssymb}
\usepackage[figuresright]{rotating}

\begin{document}

\begin{frontmatter}



\dochead{}
%% e.g. \dochead{17th International Conference on Dynamical Processes in Excited States of Solids}

\title{Robust Quantum Circuit for Clique Problem with Intermediate Qudits}


\author{Arpita Sanyal (Bhaduri)$^{1+}$, Amit Saha$^{1, 3*}$, Banani Saha$^2$, Amlan Chakrabarti$^1$}

\address{$^1$A. K. Choudhury School of Information Technology, University of Calcutta\\
$^2$Computer Science and Engineering, University of Calcutta\\
$^3$Atos, Pune, India\\~\\
$^+$arpitabhaduri@gmail.com\\
$^*$abamitsaha@gmail.com}

\begin{abstract}
Clique problem has a wide range of applications due to its pattern matching ability. There are various formulation of clique problem like $k$-clique problem, maximum clique problem, etc. The $k$-Clique problem, determines whether an arbitrary network has a clique or not whereas maximum clique problem finds the largest clique in a graph. It is already exhibited in the literature that the $k$-clique or maximum clique problem (NP-problem) can be solved in an asymptotically faster manner by using quantum algorithms as compared to the conventional computing. Quantum computing with higher dimensions is gaining popularity due to its large storage capacity and computation power. In this article, we have shown an improved quantum circuit implementation for the $k$-clique problem and maximum clique problem (MCP) with the help of higher-dimensional intermediate temporary qudits for the first time to the best of our knowledge. The cost of state-of-the-art quantum circuit for $k$-clique problem is colossal due to a huge number of $n$-qubit Toffoli gates. We have exhibited an improved  cost and depth over the circuit by applying a generalized $n$-qubit Toffoli gate decomposition with intermediate ququarts (4-dimensional qudits).

\begin{comment}

This generalized $n$-qubit Toffoli gate decomposition technique is applied using higher-dimensional qudits to attain a logarithmic depth decomposition of Toffoli gate without ancilla qudit. The circuit has been desinged for any d-ary quantum system, where $d >> 2$ with the proposed $n$-qudit Toffoli gate to obtain optimized depth of the circuit.
\end{comment}

\end{abstract}


\end{frontmatter}

% \linenumbers

\section{Introduction}
The theory of quantum computing uses quantum mechanical effects such as quantum entanglement, superposition for research. Since the early 1990s, researchers around the globe are trying to build large-scale quantum computer, as these computers have been proved to be more powerful than conventional computers, including computationally NP-problems \cite{Guo_2001, Guo_2002}. Several quantum algorithms, for example, Shor’s Algorithm \cite{31} for factoring integers, Grover’s Algorithm \cite{32} for searching an unstructured database, Triangle finding by Magniez et al.\cite{33}, Matrix Product Verification\cite{34}, graph coloring \cite{amitconf} have already been proposed and shown asymptotic improvements than their classical counterparts. 

While quantum computation is typically expressed as a two-level binary abstraction of qubits, the underlying physics of quantum systems are not intrinsically binary. In quantum computation, we work with qubits, which are 2-level quantum systems. However, it is also possible to define quantum computation with higher dimensional systems. A qu-d-it is a generalization of a qubit to a d-level or d-dimensional system. Qudits with known values for d have specific names, as an instance, a qubit has dimension 2, a qutrit has dimension 3, a ququart has dimension 4, and so on \cite{Luo2014}. In this article, the binary circuit implementation of clique problem \cite{sota, arpita}, has been addressed with temporary intermediate ququarts. The main goal of this article  is to provide an end-to-end framework for automatically implementing a clique problem using intemediate ququarts, so that anyone who can map their computational problem to the clique problem in polynomial time  and can implement it further, even if they have no prior knowledge of gate-based quantum circuit implementation.

A clique is a complete subgraph of an undirected graph in which every distinct vertex in the subgraph is connected to every other vertex through an edge. The $k$-clique problem (NP-Complete problem \cite{1, 2}) is a subset of the clique problem that asks if an arbitrary graph has a clique of size $k$. A maximum clique, on the other hand, is a complete subgraph of a graph with the biggest
size among all other complete subgraphs in the graph. Clique problems has vast range of application like pattern recognition \cite{10}, computer vision, computer analysis of financial networks \cite{12}, information retrieval \cite{11}, and spatial data mining \cite{13}. Quantum circuit design for the $k$-Clique problem has also been demonstrated in the literature \cite{sota}. In the article  \cite{sota} the circuit has been implemented using Toffoli gates. The depth and cost of the circuit depends on the multi-controlled Toffoli gates (which is decomposed into 1 or 2-qubit gates) used in the circuit. Hence, the circuit cost of the article  \cite{sota} is very high. Hence we have applied a novel $n$-qubit Toffoli decomposition technique\cite{PhysRevA.105.062453} for the optimization of the circuit in this work. A generalized $n$-qubit Toffoli gate is decomposed into two-ququart gates with logarithmic depth without using any ancilla qudit. We have also extended the approach for finding maximum clique problem \cite{21,22,23,24}.

The main contribution of the article is summarized below:

\begin{itemize}
   
    \item We exhibit a first of its kind approach to implement clique problem ($k$-clique and maximum clique) using a novel decomposition of $n$-qubit Toffoli gate using intermediate ququarts.
    \item Our approach of solving $k$-clique problem outperforms the state-of-the-art approach with respect to gate cost and circuit depth.
    \item We further extend the intermediate ququart approach to solve the maximum clique problem.
    
\end{itemize}

The structure of this article is as follows. Section 2 discusses the necessary background study to accomplish this proposed work. Section 3 proposes circuit construction for clique problem using intermediate ququarts. Section 4 analysis the efficiency of the proposed approach as compared to the state-of-the-art. Section 5 captures our conclusions.

\section{Preliminaries}
Some background studies have been discussed next.

\subsection{Clique Problem}
\begin{figure}[ht!]
\centering
\includegraphics[width=50mm, height=2cm]{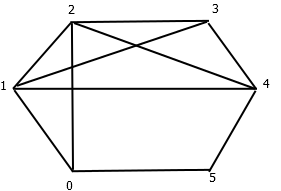}
\caption{ A graph with maximum clique }
\label{subg}
\end{figure}

A clique is a sub graph of a graph, in which all nodes are connected to each other. Particularly, if there is a sub graph of $k$ vertices that are connected to each other, we state that the graph contains a $k$-clique. Let $G(V;E)$ be a graph, where $V$ be the set of vertices and $E$ be the set of edges. A clique of a graph $G$ is a set of vertices C in which $u,v \in C \implies u,v \in E$. We say that the graph contains a $k$-clique if there is a subset of $k$ vertices that are connected to each other. A maximum clique is a complete subgraph of a graph $G$, whose size is largest among all other complete
subgraphs in G. For example if we look into the Figure \ref{subg}, we can observe that it has many cliques like $(012), (123), (124),(234), (1234)$. maximum clique of the graph is $(1234)$

\begin{figure}[ht!]
\centering
\includegraphics[width=50mm,height=3.0cm]{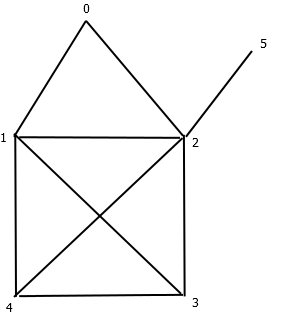}
\caption{A 6 node undirected graph  }
\label{submatch}
\end{figure}

For example Figure  \ref{submatch} represents a graph($G$) of six vertices, The $k$-clique problem asks us to determine if the graph $G$ contains the the clique of size $k$ or not, and if it does, output the vertices forming the clique. If $k=4$ the output will be {1234}, If $k=3$ output will be $(0,1,2), (1,2,4), (2,3,4)$. 
\subsection{Grover's Algorithm}
\begin{figure}[ht!]
\centering
\includegraphics[width=80mm,height=4.0cm]{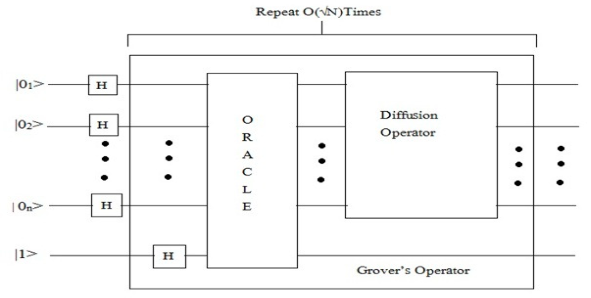}
\caption{Generalized circuit for Grover's algorithm}
\label{grover}
\end{figure}

 Grover's algorithm\cite{32} is based on amplitude amplification of the basis state which finds an element in an unsorted list. It runs in time O$(\sqrt{N})$, where $N$ is the number of elements in the list. The generalized structure of Grover's algorithm is shown in Figure \ref{grover}.

\par More elaborately, The steps of the Grover's algorithm are as follows:
 
\par \textbf{Initialization}: The first step of the algorithm is state preparation. the initial state is prepared in an equal superposition over the entire Hilbert space using Hadamard gate (H). There are other methods too for initializing the input qubits like, Dicke state or W state that will be discussed next.

\begin{figure}[ht!]
\centering
\includegraphics[width=100mm,height=3.0cm]{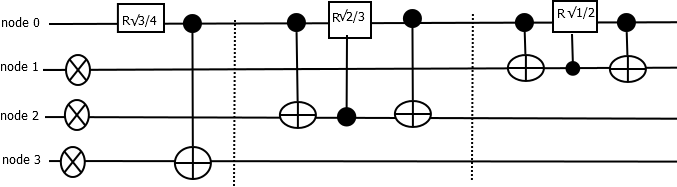}
\caption{Circuit for $\ket{D^4_3}$}
\label{dicke}
\end{figure}

\begin{figure}[ht!]
\centering
\includegraphics[scale=0.45]{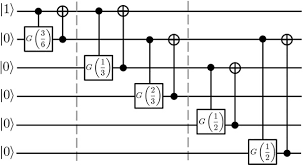}
\caption{circuit for ${W_6}$}
\label{wstate}
\end{figure}

\paragraph{\textbf{Dicke State:}} Dicke state \cite{PhysRevLett.103.020503} is a quantum state preparation technique that limits the total Hilbert spaces by using some hammimg weight. A $n$-qubit quantum state $\ket{\psi_n}$ can be expressed as the superposition of $2^n$ orthonormal basis states. $\ket{D^n_k}$ is the $n$-qubit state, which is the equal superposition state of all $n \choose k$ basis states of weight $k$. For example,$\ket{D^3_1}= \sqrt{\frac{1}{3}}(\ket{001} + \ket{010}+ \ket{100})$. Dicke state needs only polynomial number of elementary gates to prepare and have $n \choose k$ active basis states, which can be exponential in $n$ when $k = \theta(n)$. The construction of Dicke state $\ket{D^4_3}$ \cite{10.1007/978-3-030-25027-0_9} is portrayed in Figure \ref{dicke}. 
\begin{comment}

\begin{figure}[ht!]
\centering
\includegraphics[width=150mm,height=3.0cm]{fourstates.png}
\caption{Circuit for $\ket{D^6_4}$ }
\label{dicke1}
\end{figure}
\end{comment}
\paragraph{\textbf{W State:}} W state is a special case of the Dicke state. W state is a Dicke state with Hamming weight 1, such as $\ket{W} = \sqrt{\frac{1}{n}}(\ket{100...0}+···+ \ket{01...0}+\ket{00...01})$. The implementation of the W-state preparation follows from the article \cite{Cruz_2019}. The circuit size and depth of W state preparation is less than Dicke state. The circuit for the W state preparation circuit for 6 qubits is shown in Figure \ref{wstate}. The output of this circuit is the superposition of the six states $\ket{100000} + \ket{010000} + \ket{001000} + \ket{000100} + \ket{000010}+ \ket{000001}$. Limitation of W state is that it only works when clique size in $n-1$.

\textbf{Oracle}: Grover’s algorithm works with a unitary operator $O$ called the oracle function, that add a negative phase to the solution states. The functional view of Grover's Search Algorithm is presented here. 
 $\exists$ function oracle $(O)$ such that

\[
   						 O\ket{x}= 
							\begin{cases}
    							-\ket{x},& \text{if \textit{x} is Marked}\\
                                 \ket{x},              & \text{otherwise}
\end{cases}
\]
This oracle will be a diagonal matrix, where the entry that correspond to the marked item will have a negative phase.

\textbf{Amplitude Amplification}: We need to perform the inversion about the average of all amplitudes of quantum state for an optimal number of iterations to make the amplitude of the marked state large enough so that it can be obtained from a measurement with probability close to 1.  

\textbf{Number of Iterations}: Iterations of Grover’s algorithm are the number of times that the oracle and amplification stages are performed.
Each iteration of the algorithm increases the amplitude of the marked state by $O(\sqrt{1/N})$. Grover's search algorithm requires $\sqrt{N/M}$  iterations to get the probability of one of the marked states $M$ out of total $N$ number of states set.
Therefore, optimal number of iterations are the key for getting proper result. 
\par{\textbf{Diffusion operator}}:
This procedure amplifies the amplitude of the marked item, and shrinks the other item's amplitudes, so that measuring the final state will return the right item with near certainty. The generalized matrix representation of the diffusion operator is shown in Table \ref{diffusion}. A six-qubit diffusion operator is also presented in Table \ref{diffusion}. As shown in six-qubit diffusion operator, a 6-qubit Toffoli gate is required. Therefore, for $n$-qubit diffusion operator, $n$-qubit Toffoli gate is needed. This $n$-qubit Toffoli gate needs to be realized into one-qubit or two-qubit gates. While decomposing the $n$-qubit Toffoli gate, if the depth and the ancilla qubits increase arbitrarily then the run-time complexity of the algorithm also increases, which is undesirable.
\begin{table}[htb]
\centering
\caption{Diffusion Operator}
%\scriptsize
\begin{tabular}{ c | c }
\hline
 Six-qubit Diffusion Circuit & Generalized Matrix Representation \\ \hline
\\ \includegraphics[width=3cm]{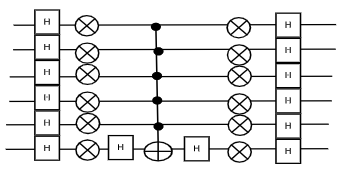} &
$\begin{pmatrix}
 -1 + \frac{2}{N} & \frac{2}{N} & \ldots & \frac{2}{N}\\ 
 \frac{2}{N} & -1 + \frac{2}{N} & \ldots & \frac{2}{N}\\
\vdots & \vdots  & \ddots & \vdots\\ 
\frac{2}{N} & \frac{2}{N} & \ldots &  -1 + \frac{2}{N}
\end{pmatrix}$ \\
\\ \hline
\end{tabular}
\label{diffusion}
\end{table}

\subsection{Multi-controlled Toffoli Decomposition with Intermediate Qudits}

This section starts with the discussion of qudit technology followed by Toffoli decomposition. \cite{PhysRevA.105.062453}

\subsubsection{Qudit}
Qudit technology \cite{Gokhale_2019} is emerging as an alternative  for quantum computation and quantum information science. Qudits provide a larger state space to store and process information and can do multiple control operations simultaneously. These features reduces the circuit complexity. Traditionally, quantum computation is expressed in terms of quantum bits, or qubits. We designate these two basis states as $\ket{0}$ and $\ket{1}$ and can represent any qubit as \\
$\psi=\alpha \ket{0} + \beta\ket{1}$ \\ with $|\alpha^2 + \beta^2 |=1$.
 A qudit is the unit of quantum information for d-ary quantum systems. Qudit states can be expressed by a vector in the d-dimensional Hilbert space $H_d$. The vector space is the span of orthonormal basis vectors $\ket{0}, \ket{1},\ket{2} ,\dots \ket{d-1}$. N qubits can be expressed as $\frac{N}{\log_2 d}$ qudits. This reduces runtime of the algorithm by a $\log_2d$ factor. Qudit state can be described as \\
$\psi=\alpha_0\ket{0}+ \alpha_1\ket{1} + \alpha_2\ket{2}+ \dots \alpha_{d-1}\ket{d-1}$.\\ The state of a qudit is transformed by qudit gates Some useful gates of qudit systems are discussed next.\cite{Muthukrish-2000}

\paragraph{Generalized CNOT gate} For d-ary quantum
systems, the binary two-qubit CNOT gate is generalized
to \\ $C_{X,d}\ket{x}\ket{y}=\ket{x}\ket{(y + 1) \mod d}$ \\ only if $x=d-1$ and
otherwise = $\ket{x}\ket{y}$.
\paragraph{Generalized $n$-qudit Toffoli gate} We extend the generalized CNOT further to operate over $n$ qudits as a generalized $n$-qudit Toffoli gate $C^n_{X,d}$. For $C^n_{X,d}$ , the target qudit is incremented by 1 mod donly when all the $n-1$ control qudits are $d-1$.

\begin{comment}

The $(d^n × d^n )$ matrix representation of the generalized $n$-qudit Toffoli gate is as follows:\par
$C^n_{X,d}$ = $\begin{pmatrix} 
\\ I_d & 0_d & 0_d \dots 0_d
\\ 0_d & I_d & 0_d \dots 0_d
\\ 0_d & 0_d & I_d \dots 0_d
\\ \dots & \dots & \dots
\\ 0_d & 0_d & 0_d \dots X_d
\end{pmatrix}$ \\
where $I_d$ and $0_d$ are both $d \times d$ matrices as shown below:\par
$I_d$ = $\begin{pmatrix} 
\\ 1 & 0 & 0 \dots 0
\\ 0 & 1 & 0 \dots 0
\\ 0 & 0 & 1 \dots 0
\\ \dots & \dots & \dots
\\ 0 & 0 & 0 \dots 1
\end{pmatrix}$ \\ \par
$0_d$ = $\begin{pmatrix} 
\\ 0 & 0 & 0 \dots 0
\\ 0 & 0 & 0 \dots 0
\\ 0 & 0 & 0 \dots 0
\\ \dots & \dots & \dots
\\ 0 & 0 & 0 \dots 0
\end{pmatrix}$ \\

\end{comment}
\subsubsection{n-qubit Toffoli Decomposition}
\begin{figure}[ht!]
\centering
\includegraphics[scale=0.3]{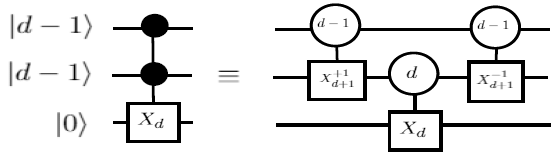}
\caption{Generalized Toffoli in d-ary quantum systems}
\label{gentoffoli}
\end{figure}

Toffoli gate is the central building block of several quantum algorithms.  Since the Toffoli involves 3-body interactions, it cannot be implemented naturally in a real quantum devices. Toffoli gate can be constructed by decomposing it into single and two qubit gates. For example CNOT gates require 6 such gates plus 10 single qubit gates;\cite{35}  For generalized Toffoli gates the resources increase rapidly, requiring $O(n^2)$ two-qubit gates. Toffoli gates can be constructed efficiently using fewer resources than previous designs with the help of intermediate qudits \cite{Gokhale_2019}. \\
The Figure \ref{gentoffoli} shows the docomposition of generalized Toffoli gate \cite{PhysRevA.105.062453} in any finite dimensional quantum system or d-ary quantum systems. Target qudit is incremented ($X_d$ operation) when two-controlled qudits are both $\ket{d-1}$. First, a $\ket{d-1}$ controlled $X^{+1}_{d+1}$ operation increment the target qudit by $1 \ mod \ (d+1) $. This changes the value of the second qudit to $\ket{d}$ if and only if the first and the second qudits were both $\ket{d-1}$. Then, a $\ket{d}$  controlled $X_d$ gate is applied to the target qudit. Therefore, $X_d$ is executed only
when both the first and the second qudits are $\ket{d-1}$. The controls are reinstated to their original states by a $\ket{d-1}$ controlled $X^{-1}_{d+1}$ gate, which reverses the effect of the first gate. Next we shall apply this technique to decompose the 2-controlled, 3-controlled, 4-controlled and 5-controlled Toffoli gates.

\begin{figure}[ht]
\centering
\includegraphics[width=80mm,height=3.0cm]{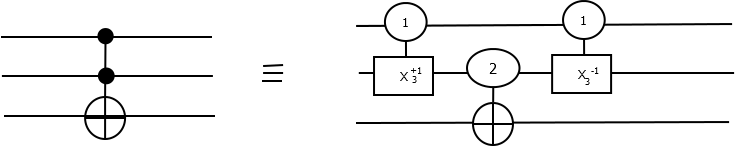}
\caption{2-controlled Toffoli Gate Decomposition }
\label{2ctrl}
\end{figure}

\begin{figure}[ht!]
\centering
\includegraphics[width=100mm,height=3.0cm]{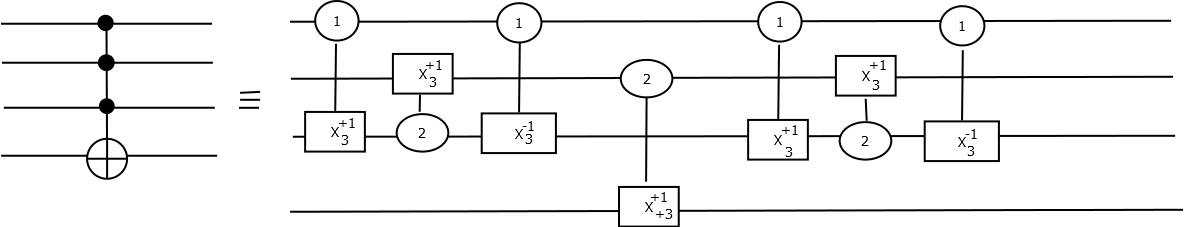}
\caption{3-controlled Toffoli Gate Decomposition }
\label{3ctrl}
\end{figure}

\begin{figure}[ht!]
\centering
\includegraphics[width=80mm,height=3.0cm]{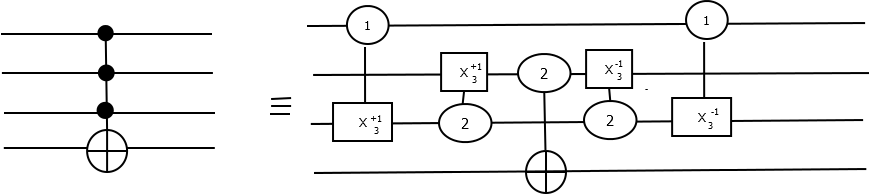}
\caption{3-controlled Toffoli Gate Optimized Decomposition }
\label{3ctrlopt}
\end{figure}

\begin{figure}[ht!]
\centering
\includegraphics[width=80mm,height=3.0cm]{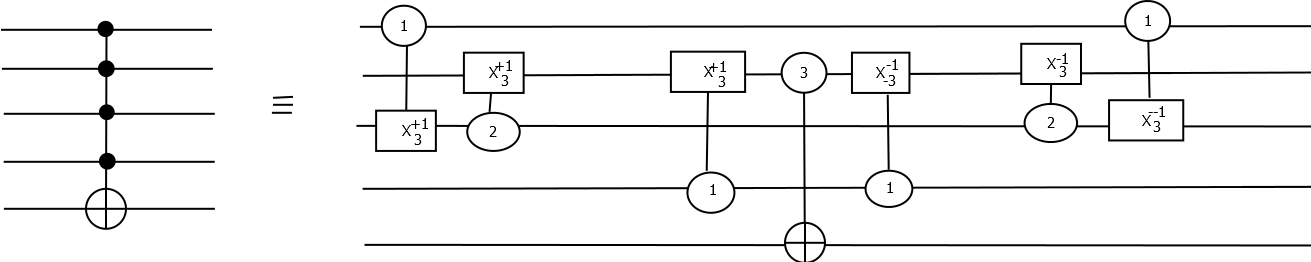}
\caption{4-controlled Toffoli Gate Optimized Decomposition }
\label{4ctrlopt}
\end{figure}

\begin{figure}[ht!]
\centering
\includegraphics[width=100mm,height=3.0cm]{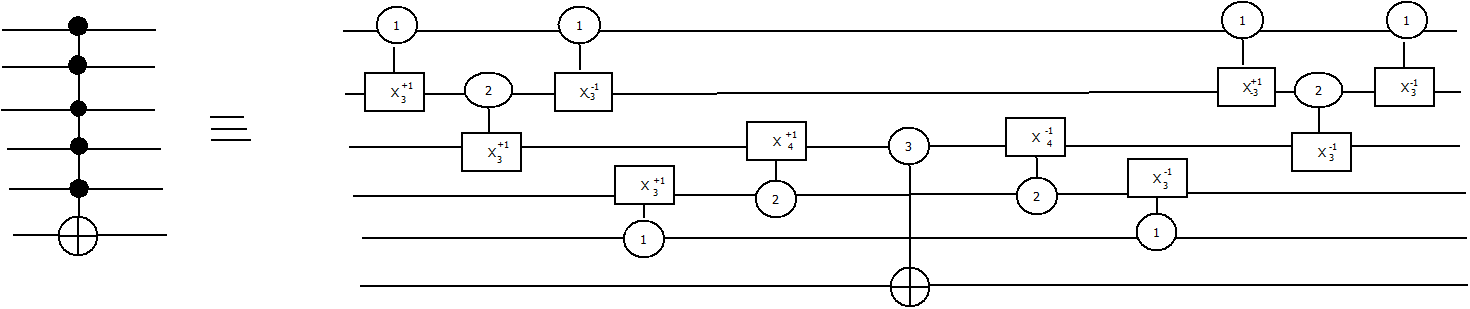}
\caption{5-controlled Toffoli Gate Decomposition }
\label{5ctrl}
\end{figure}

\begin{figure}[ht!]
\centering
\includegraphics[width=100mm,height=3.0cm]{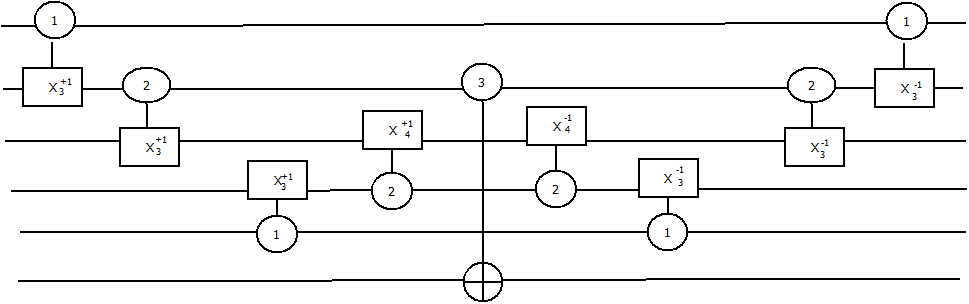}
\caption{5-controlled Toffoli Gate Optimized Decomposition }
\label{5ctrlopt}
\end{figure}

The generalized Toffoli gate decomposition as shown in the Figure \ref{gentoffoli} has been used for the decomposition of 2-controlled, 3-controlled, 4-controlled and 5-controlled Toffoli gates, which are used as examples for the circuit construction of clique problem in subsequent sections. The Figure \ref{2ctrl} shows the decomposition of a 2-controlled Toffoli gate. Second qubit is incremented if the first control qubit is 1. $x^{+1}_3$ is modulo 3 increment gate, that increments the value of the second qubit. Finally last qubit is flipped is the control qubit is 2. The qubits values are uncomputed to their initial value.
\\ The Figure \ref{3ctrl} shows the decomposition of a 3-controlled Toffoli gate. Third qubit is incremented if the first control qubit is 1. $x^{+1}_3$ is modulo 3 increment gate, Second qubit is incremented if third qubit is 2. The target qubit is incremented if 2nd qubit is 2. The qubits values are uncomputed to their initial value.
\\Further, the optimized 3-control Toffoli gate decomposition is given in Figure  \ref{3ctrlopt}, where two generalized CNOT gates operating one after another on the same qudits are removed by applying the optimization rule as described in \cite{38}
\\ Decomposition of 4-controlled Toffoli gate is shown in Figure \ref{4ctrlopt}. 
\\Decomposition of 5-controlled Toffoli gate is shown in Figure  \ref{5ctrl}. The decomposition is done in the similar manner.
\\The optimized decomposition is shown in the Figure \ref{5ctrlopt}. As shown in Figure \ref{5ctrlopt}, to decompose 5-controlled Toffoli, we need to temporarily access 4-dimensional quantum systems or ququarts. In fact, in similar way for $n>4$, $n$-qubit Toffoli can be decomposed using in intermediate ququarts.
\section{Quantum Circuit Construction for $k$-clique problem with Intermediate Qudits} Implementation of $k$-clique problem follows the method used in the article of Metwalli et al. \cite{sota}. For finding $k$-clique of a graph using Grover's algorithm, we have to go through three steps namely input state preparation, oracle construction and amplification of the marked state. For the amplification of the marked states, we use Grover's diffusion operator. Since Diffusion operator is problem independent, therefore we shall directly apply the operator as we have already exhibited in the Section 2. Therefore, only the first two steps have been detailed in the subsequent subsections.

\subsection{Input State Preparation}
In Grover’s algorithm, the states are prepared in an equal superposition of the whole Hilbert space using the H (Hadamard) gate. Initializing into full superposition needs only $n$ H gates and time complexity $O(1)$, since all H gates can be run simultaneously. Albeit, it unnecessarily searches the full Hilbert space. Here we use Dicke states and W state to limit the search spaces. If we want to search for a clique size four as shown in Figure \ref{submatch}, we shall limit the search space to only the subgraphs containing four vertices. The Dicke state preparation of the 4 nodes from 6 nodes is shown in Figure \ref{dicke}. After applying Dicke state circuit the input states will be the superposition of \\
$\ket{\psi}= \frac{1}{\sqrt{2}} \ket{011011} + \ket{011110} + \ket{001111} + \ket{011101} +\ket {111010} + \ket{111100} + \ket{010111} +\ket{101101}+ \ket{110011} + \ket{101011} +
\ket{110110} + \ket{101110} + \ket{111001} + \ket{110101} + \ket{100111}$.\\

\subsection{Oracle Construction for $k$-clique}
For finding a clique in a given graph, we have to check the total number of edges and vertices.
Each node of the graph is represented by a qubit and the edges between them is checked using one or more multi-controlled Toffoli ($n$-qubit Toffoli) gates connecting respective qubits. Number of nodes are also tested using one or more multiple-controlled Toffoli ($n$-qubit Toffoli) gates. For the physical realization of Toffoli gate, it is decomposed into several 1-qubit or 2-qubit gates. Hence, the increasing number of Toffoli gates increase the circuit depth and size. In this work, we try to decompose these $n$-qubit Toffoli gates with intermediate higher dimensional qudits (qutrits or ququarts) for better circuit robustness.

There are two variations of this oracle one is Checking-based and another is Increment-based as discussed in the article \cite{sota}. Each oracle is detailed in the subsequent subsections.
\subsubsection{Checking-based Oracle}
In Checking-based oracle, there is an edge counter and a node counter. Each $n$-qubit Toffoli gates connecting the qubits represent one edge. When an edge is encountered, the value of the edge counter is increased by one. After all edges have been counted, the results are checked. If we want to find the clique of size 4 in graph $G$ as in Figure \ref{submatch}, then after applying $n$-qubit Toffoli gates, we need to check that we have precisely six edges (binary equivalent is $110_2$). We need $\log{k \choose 2}$ qubits to represent the edge counter. We require a 3 qubit counter here that can count up to 7 that is ($111_2$). We need an edge flag to check the value of the edge counter. Similar process is used to count the nodes using node counter. If the node counter contains the correct value then we may claim that a clique of size $k$ is found. The following Figure \ref{checkedge}, \ref{checknode} and \ref{targetflip} shows Checking-based oracle of the graph of Figure \ref{submatch}. The graph $G$ has nine edges as well as six nodes. The Figure \ref{checkedge} shows the oracle for checking the edges of a complete sub-graph of size 4. The Figure \ref{checknode} shows the oracle for checking the node count. Three ancilla qubits are used here for counting six edges and four nodes.

\begin{figure}[ht!]
\centering
\includegraphics[scale=.25]{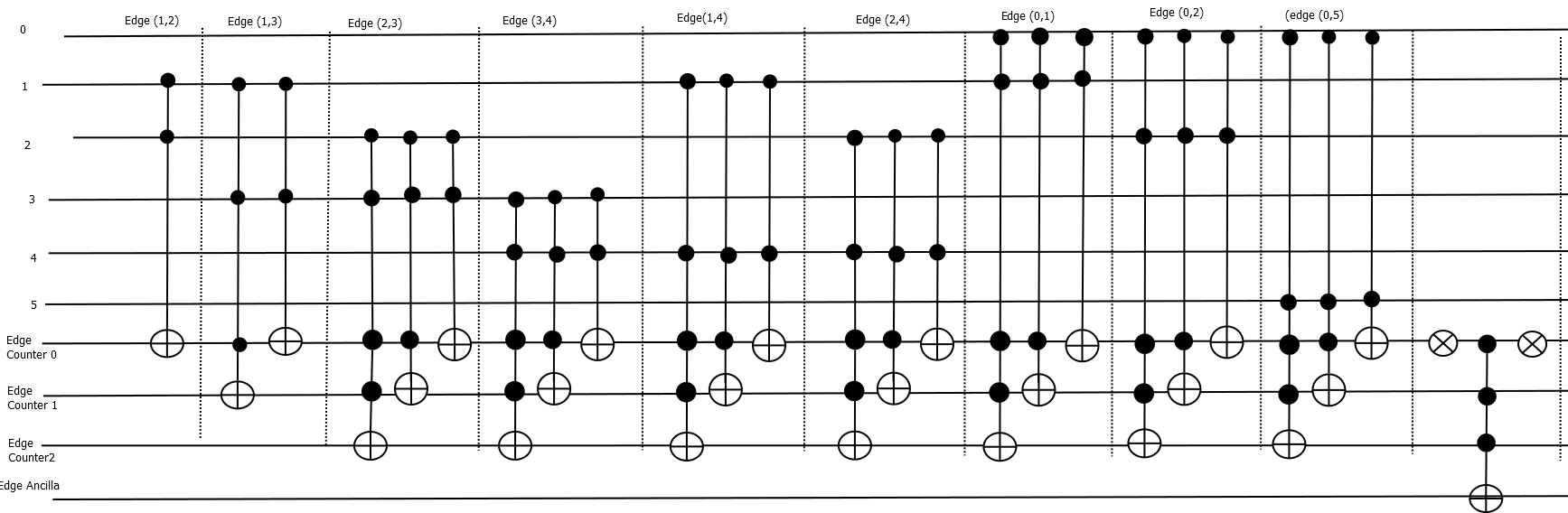}
\caption{Checking-based edge oracle }
\label{checkedge}
\end{figure}

\begin{figure}[ht!]
\centering
\includegraphics[scale=0.4]{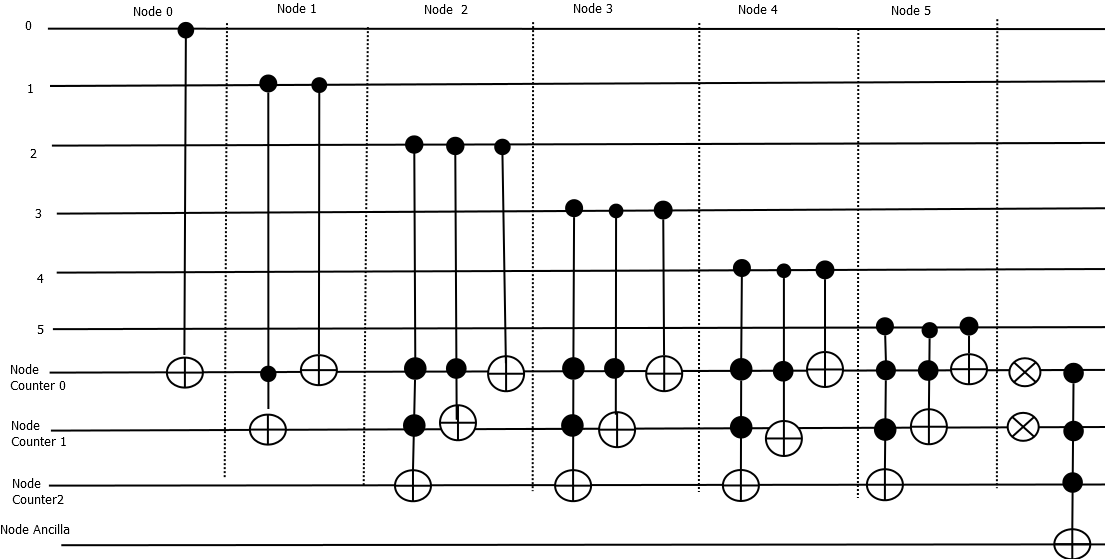}
\caption{Checking-based node oracle }
\label{checknode}
\end{figure}

If both the node flag and edge flag are high, then the target qubit is flipped. Figure \ref{targetflip} shows the whole procedure.

\begin{figure}[ht!]
\centering
\includegraphics[scale=0.5]{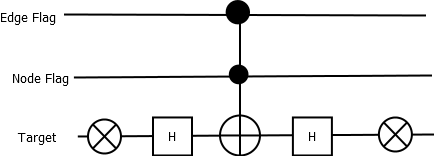}
\caption{Checking-based node oracle }
\label{targetflip}
\end{figure}

\subsubsection{Incremental-based Oracle}
For incremental-based oracle, an increment operator is applied. There is an one qubit edge flag. The edge flag becomes 1 if and only if an edge exists between two nodes. Edge flag is then used as a control qubit for an increment circuit that adds 1 every time it finds an edge. The following Figure \ref{incredge} shows Incremental-based oracle of the graph of Figure \ref{submatch}. The graph $G$ has a sub-graph of six edges as well as four nodes. Hence, it contains a clique of size 4.

\begin{figure}[ht!]
\centering
\includegraphics[scale=.25]{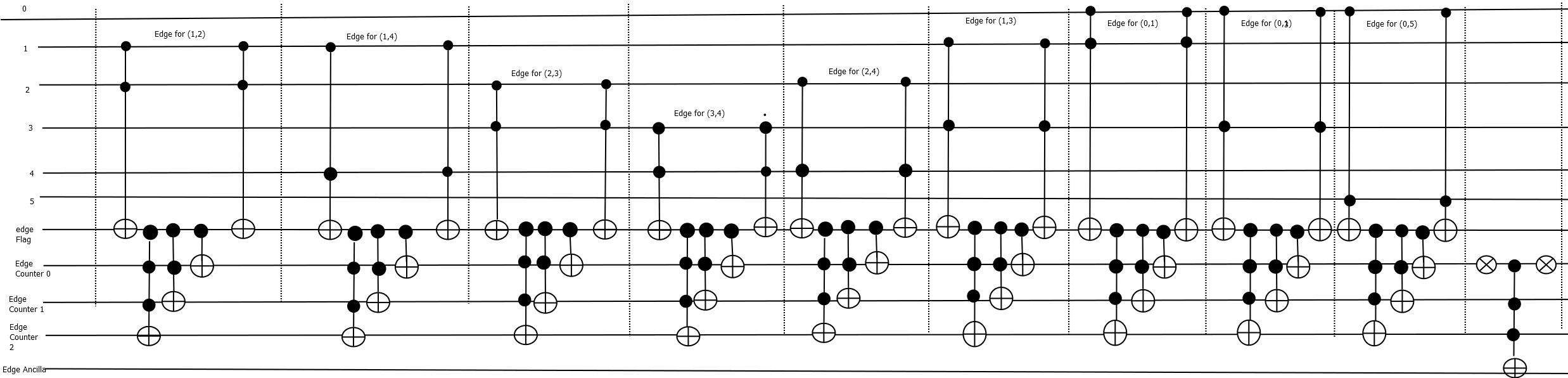}
\caption{Increment-based edge oracle }
\label{incredge}
\end{figure}
\begin{figure}[ht!]
\centering
\includegraphics[scale=0.35]{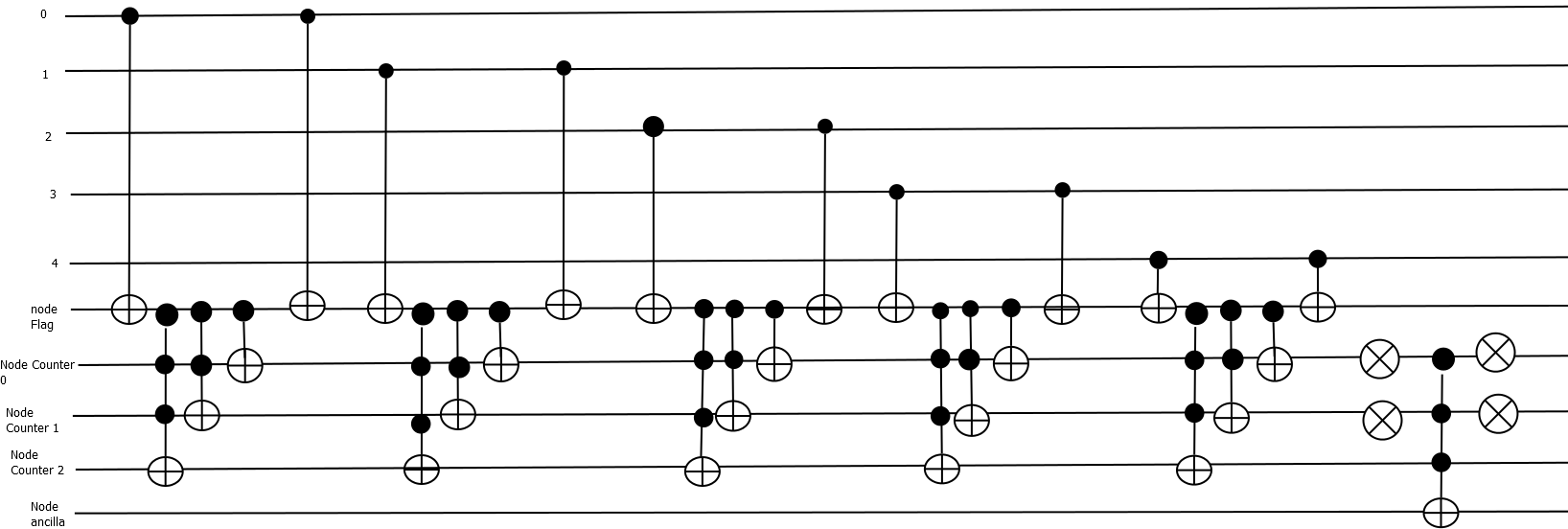}
\caption{Increment-based node oracle }
\label{incrnode}
\end{figure}

The Figure \ref{incredge} shows the Incremental-based oracle for counting six edges. The Figure \ref{incrnode} shows the Increment-based oracle for counting four nodes. The size of both the node counter and edge counter is 3.

If the both node flag and edge flag are high then the target qubit is flipped as we can observe in the Figure \ref{targetflip}. As we can observe from the Figure \ref{checkedge}, \ref{checknode}, \ref{incredge} and \ref{incrnode}, a significant number of Toffoli gates has been used for the construction of these oracular circuits. Implementation of these large number of Toffoli gates increases circuit depth accordingly as detailed in the article \cite{sota}, which makes the circuit impossible to implement in the real devices. Hence, the Toffoli gates are decomposed using intermediate qudit gates as discussed in section 2. This decomposition technique using qudit gates reduces the size and depth of the circuit remarkably as shown in the Figure \ref{2ctrl}, \ref{3ctrlopt}, \ref{5ctrlopt}.

\paragraph{Oracle Construction for $k$-clique Problem}
The automated and generalized algorithm of quantum circuit synthesis for $k$-clique problem is discussed here. $k$-clique problem finds a clique of size $k$ in a graph.
The steps of the circuit synthesis for $k$-clique are shown below: \\
  \textbf{Input} : {Graph $G(V,E)$, Size of the clique to be searched that is $k$} \\
  \textbf{Output}: {Clique of size $k$}
\begin{itemize}
  \item \textbf{Step 1}: The total number of vertices are the input qubits of the circuit. Input states are prepared using Dicke state, W state or Hadamard gates.
  \item \textbf{Step 2}: Increment-based oracle or Checking-based oracle is used to count the number of nodes and edges of a sub-graph of size $k$. Edge counter should have the value $k \choose 2$ and node counter should have the value $k$. An edge flag is used to detect the correct value of the edge counter and a node flag is used to detect the correct value of the node counter.
  \item \textbf{Step 3}: If both the edge flag and node flag matches then a clique of size $k$ is found.
  
\end{itemize}

\paragraph{Oracle Construction for Maximum Clique Problem}
The automated and generalized algorithm of quantum circuit synthesis for maximum clique problem is discussed here. maximum clique problem finds clique in a graph with maximum cardinality. The algorithm starts with finding clique of size $k$ = $n$ ($n$ is the total number of vertices of the graph) using the oracle of $k$-clique. If a clique is found, it is the maximum clique of the graph. Otherwise, a clique is searched for size $n-1$. The process continues up to $n = 2$. \\
The steps of the circuit synthesis for MCP are shown below:\\
 \textbf{Input} : {Graph $G(V,E)$} \\
  \textbf{Output}: {Maximum Clique of the graph} 
\begin{itemize}
  \item \textbf{Step 1}: Algorithm for finding $k$-clique is used for searching clique of size $n$, where the number vertices of the given graph are $k$.

  \item \textbf{Step 2}: If a clique is found, then exit from the process. 
  \item \textbf{Step 3}: If clique is not found, then clique size $n$ is reduced to $n-1$ and repeat the whole process from step 1.
  \item \textbf{Step 4}: At each iteration, the value of $n$ is reduced by 1 $(n, n-1,n-2,n-3,\dots, 2)$ until a clique is found.

\end{itemize} 

If we take a look at the Figure \ref{submatch}, there are 6 nodes and 9 edges. The maximum clique problem starts with $k=6$. A complete graph of 6 nodes contains $6 \choose 2$, 15 edges. But the graph does not have 15 edges. Hence, next we try to find the clique of size $k=5$. A complete graph of 5 nodes contains $5 \choose 2$, that means 10 edges. But the graph has only 9 edges. Hence, we move to the next iteration of searching with the clique size $k=4$. We find the total edge as 6 for 4 vertices in a subgraph. We stop the searching process and conclude that the maximum clique size of the graph is 4 of Figure \ref{submatch}.

\section{Implementation Results and Analysis}
 The behaviour of a quantum algorithm can be realized by a simulator on a classical computer. Simulator sense how the quantum circuit may be implemented on a real quantum computer. The circuit for $k$-clique problem is simulated on Google Colaboratory platform \cite{Bisong2019}.
 
\subsection{Simulation with full Hilbert space}
The oracle circuit for $k$-clique problem for the graph shown in Figure \ref{submatch} has been implemented. Here, we have 6 qubit lines for inputs, there can be a total of $2^6=64$ different qubit combinations as $\ket{000000}$, $\ket{000001}$, $\dots$, $\ket{111111}$ after applying Hadamard gate to all input qubits. Hence, the database on which Grover's algorithm is applied contains 64 different elements.
\subsubsection{Simulation using Checking-based oracle}
\begin{figure}[ht!]
\centering
\includegraphics[width=150mm,height=3.0cm]{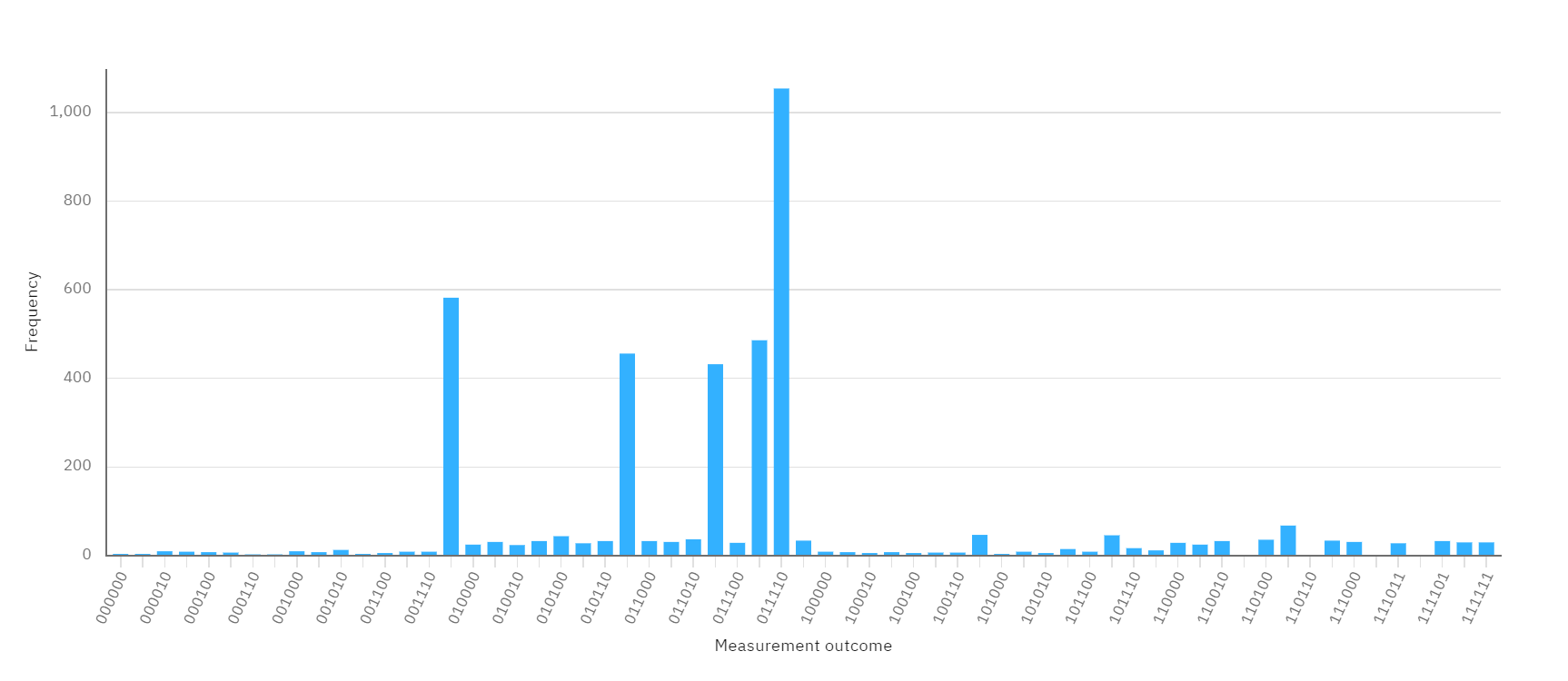}
\caption{Ideal simulation of Checking-based oracle}
\label{Checkingop}
\end{figure}

Figure \ref{checkedge}, \ref{checknode},  shows the complete gate-level synthesis of Grover's circuit for the 4-clique problem. It is to be noted that all the Toffoli or multi-controlled Toffoli gates are decomposed with intermediate qudits before performing the simulation procedure. The simulation steps required for the synthesis are as follows:
\begin{enumerate}
%[Step 1:]
\item The oracle checks whether four vertices are connected with six edges or not. It then inverts the amplitude of the solution state for which the four vertex combination forms a sub graph with six edges. The solution state for this case is $\ket{011110}$, 
\item The output of the oracle is acted upon by the Grover's diffusion operator as shown in Figure \ref{Checkingop}. The diffusion operator amplifies the amplitude of the marked solution state.
\item Steps 1 and 2 constitute the Grover's operator for the Grover's search algorithm. Hence, these two steps are repeated $\dfrac{\pi}{4}\sqrt{\frac{64}{1}}$ times. (For an $N$ item database with $M$ number of known solutions \cite{32}, Grover's iteration must be repeated $\dfrac{\pi}{4}\sqrt{\dfrac{N}{M}}$ times in order to obtain a solution). If our number of solutions are not previously known \cite{Yoder_2014, Long_1999, Long2002PhaseMC, Gui_Lu_2006, https://doi.org/10.1002/que2.49, Ding_2019}, therefore we need to consider a tricky method for the iterations without any asymptotic penalty. As per \cite{Boyer_1998}, we need to start with a single application of the Grover Iterate. If the marked state is not found, repeat the Grover's algorithm with a number of iterations which is $6$ times the previous number.
\end{enumerate}
The resultant output after applying Grover's operator $\dfrac{\pi}{4}\sqrt{1}$ $\sim$ $O(1)$ time is shown in Figure \ref{Checkingop}, where the amplitude of the solution state has been amplified. The location of the solution state is $\ket{011110}$, which is the vertex combination in an example graph of Figure \ref{Checkingop} that forms clique with high probability. The circuit for clique size three has been simulated. The marked state $\ket{011110}$ is in high amplitude as shown in the output Figure \ref{Checkingop}, which depicts that $1$, $2$, $3$ and $4$ form a square. Similarly, the simulation of Increment-based oracle can be performed.

\subsubsection{Triangle finding problem using Dicke state preparation}
\begin{figure}[ht!]
\centering
\includegraphics[width=30mm,height=3.0cm]{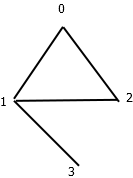}
\caption{One Triangle}
\label{onetri}
\end{figure}

\begin{figure}[ht!]
\centering
\includegraphics[width=150mm,height=3.0cm]{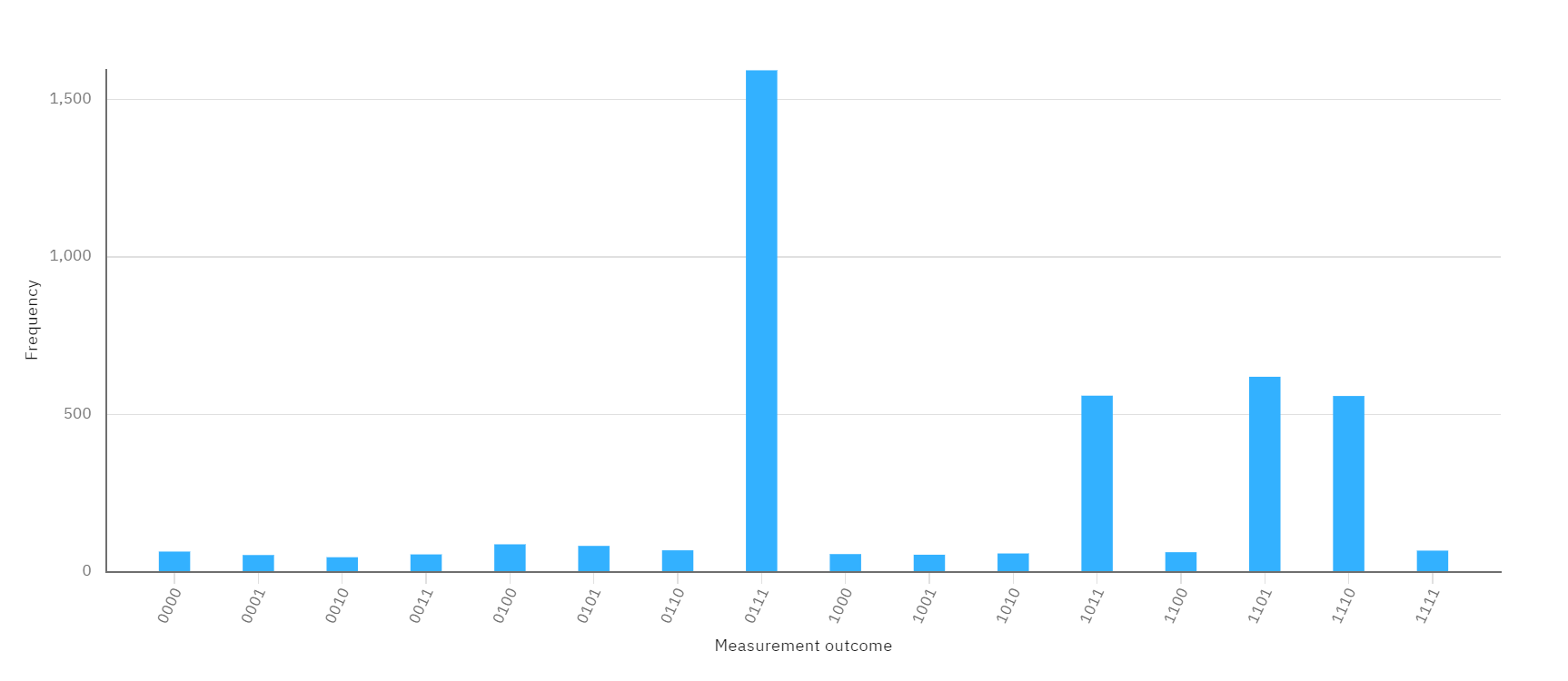}
\caption{Output of Checking-based oracle for finding triangle}
\label{incr3op}
\end{figure}

\begin{figure}[ht!]
\centering
\includegraphics[width=150mm,height=3.0cm]{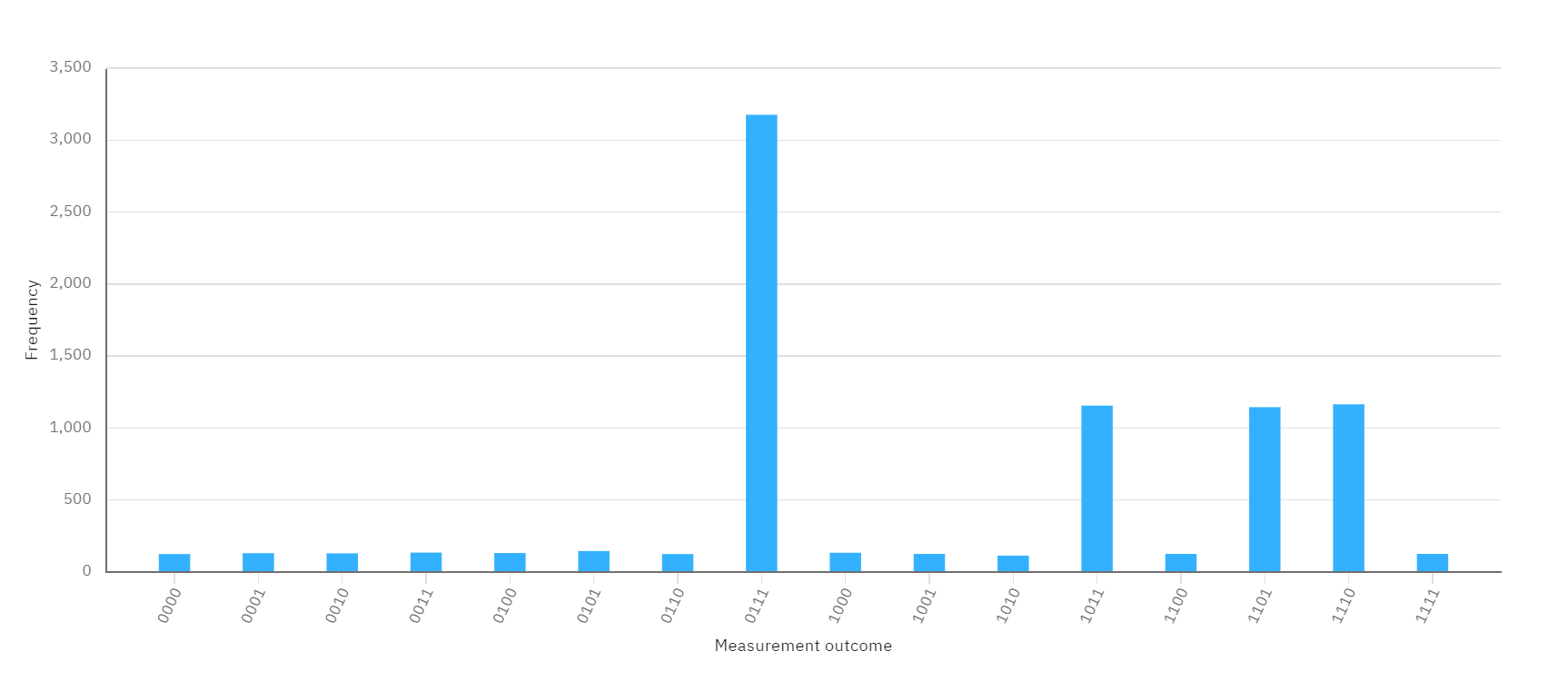}
\caption{Output of Increment-based oracle for finding triangle}
\label{check3op}
\end{figure}

In similar way, the implementation result of applying Checking-based oracle and Increment-based oracle in Google Colaboratory platform for finding a traingle in the graph in Figure \ref{onetri} is shown in the Figure \ref{incr3op} and Figure \ref{check3op}. The input state preparation is used as Dicke state for the both oracle.

\subsection{Circuit Cost Analysis}
In order to test the efficiency of our implementation, we compared various combinations of the problem variables with the state-of-the-art problem. The comparison is based on the gate count (size) and the depth of the circuit proposed in the state-of-the-art article and the gate count (size) and the depth after applying the Toffoli decomposition with intermediate qudit  technique. The circuit cost and depth have been analysed for both of the approaches i.e., Checking-based and Increment-based oracle. Before that let's discuss about the cost of multi-controlled Toffoli decomposition with intermediate qudits. 

\subsubsection{Cost of Decomposed Toffoli Gate }
\begin{figure}[h!]
    \centering
    \includegraphics[width=0.5\textwidth]{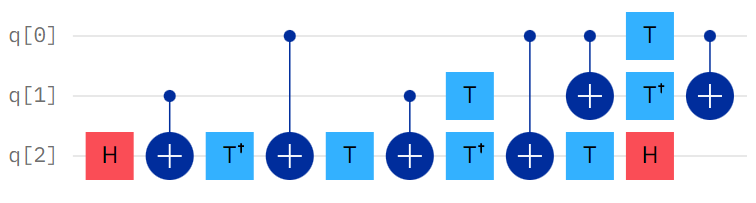}
    \caption{Decomposed Toffoli (Circuit Depth 12, T-count 7, CNOT count 6, H count 2).}
    \label{fig:tof_selinger}
\end{figure}

As discussed Multi-controlled Toffoli gates can not be directly used in real quantum devices. To map those gates, it needs to be decomposed into several 1-qubit and 2-qubit gates. The circuit depth grows with the number of Toffoli controls when the MCT gate is implemented on the IBM Q devices. The total number of elementary gates required to map one Toffoli gate is 15 and the depth is 12 as shown in Figure \ref{fig:tof_selinger}. But if we use the decomposition technique that is shown in the Figure \ref{3ctrlopt}, we can observe that the Toffoli gates can be decomposed using only three 2-qutrit gates and the depth is also three. Hence, we can observe a significant improvement in terms of the cost and depth of the circuit. Table \ref{Gate cost for Toffoli gates} shows Multi-controlled Toffoli gate variations with total number of gates needed to implement it using conventional method and using intermediate qudit. 

 Total number of 1-qubit, 2-qubit gates required to realize Toffoli gate is given in Table \ref{Gate cost for Toffoli gates}. As we observe from the Table \ref{Gate cost for Toffoli gates} that the total number of gates to map one Toffoli gate is 15. Using the Toffoli decomposition technique we can reduce it. The total number of Toffoli gate for different state preparation techniques and the corresponding cost and depth comparison of the circuit is shown in the Table \ref{checking oracle gates}. 

\begin{table}[ht!]
\centering
\caption{Toffoli gate cost comparison using qubit and qudit gates}
%\scriptsize
\begin{tabular}{| c | c | c | c | c | c |c|}
\hline
  Decomposition technique & Number of control & size & Depth & 1-qubit gates & 2-qubit gates & 2-qudit gates  \\ \hline
 Standard  \cite{sota} &2 & 15 & 12 & 9 & 6 & 0 \\ \hline
 \rowcolor{LightCyan}Intermediate Qudit & 2 & 3 & 3 & 0 & 0 & 3 \\ \hline
 Standard  \cite{sota} & 3 & 27 & 22 & 14 & 13 & 0 \\ \hline
 \rowcolor{LightCyan}Intermediate Qudit & 3 & 5 & 5 & 0 & 0 & 5 \\ \hline
 Standard \cite{sota} & 4 & 77 & 60 & 48 & 29 & 0  \\ \hline
 \rowcolor{LightCyan}Intermediate Qudit & 4 & 7 & 7 & 0 & 0 & 7 \\ \hline
 \end{tabular}
\label{Gate cost for Toffoli gates} 
\end{table}
The total number of NOT, CNOT and Toffoli gates required for finding triangle in an arbitrary graph is shown in the Table \ref{Gate cost for Toffoli gates} of the article \cite{sota}.

\begin{table}[ht!]
\centering
\caption{Toffoli count for Checking-based oracle}
%\scriptsize
\begin{tabular}{|c | c | c | c | c |}
\hline
  Method used & State Preparation& Total & Size & Depth \\ \hline
  Standard  \cite{sota} & Hilbert & 63 & 945  & 756 \\ \hline
  \rowcolor{LightCyan}Intermediate Qudit & Hilbert & 63 & 189  & 189 \\ \hline
  Standard  \cite{sota} & Dicke & 18 & 270  & 216 \\ \hline
  \rowcolor{LightCyan}Intermediate Qudit & Dicke & 18 & 54  & 54 \\ \hline
 \end{tabular}
\label{checking oracle gates} 
\end{table}
The Toffoli gate cost for Increment-based oracle for different state preparation techniques is given in the Table \ref{increment oracle gates}. 
\begin{table}[ht!]
\centering
\caption{Toffoli count for Increment-based oracle using qutrit}
%\scriptsize
\begin{tabular}{| c |c | c | c | c |}
\hline
  Method used & State Preparation & Total & Size & Depth  \\ \hline
 Standard  \cite{sota} & Hilbert & 48 & 720 & 576 \\ \hline
 \rowcolor{LightCyan}Intermediate Qudit & Hilbert & 48 & 144 & 144 \\ \hline
 Standard  \cite{sota} & W state & 8 & 120 & 96 \\ \hline
 \rowcolor{LightCyan}Intermediate Qudit & W state & 8 & 24 & 24 \\ \hline
 Standard  \cite{sota} & Dicke state & 26 & 390 & 312 \\ \hline
 \rowcolor{LightCyan}Intermediate Qudit & Dicke state & 26 & 78 & 78 \\ \hline
 
 \end{tabular}
\label{increment oracle gates} 
\end{table}
As we can observe from the Table \ref{increment oracle gates} that the size and depth of the circuit get reduced if we use higher dimensional Toffoli decomposition technique.
\begin{comment}
\begin{table}[ht!]
\centering
\caption{Gate cost for Toffoli gates}
%\scriptsize
\begin{tabular}{| c | c | c | c | c | c |}
\hline
  Full Search Space & W State Prep & Dicke State Prep & size & Depth &\\ \hline
 Standard   \cite{sota} &2 & 15 & 12 & 9 & 6 \\ \hline
 \rowcolor{LightCyan}Intermediate Qudit & 2 & 3 & 3 & 0 & 3 \\ \hline
 Standard   \cite{sota} & 3 & 27 & 22 & 14 & 13 \\ \hline
 \rowcolor{LightCyan}Intermediate Qudit & 3 & 5 & 5 & 0 & 5 \\ \hline
 Standard  \cite{sota} & 4 & 77 & 60 & 48 & 29  \\ \hline
 \rowcolor{LightCyan}Intermediate Qudit & 3 & 5 & 5 & 0 & 5 \\ \hline
 \end{tabular}
\label{Checking Based Oracle} 
\end{table}
\end{comment}
\subsubsection{Total Cost for clique problem}
Quantum circuit becomes difficult to implement on hardware if the number of gates of the circuit is very high because of its erroneous behaviour. Hence we have to find the circuit size and the depth of the circuit implementations for two kind of oracles with different state preparation techniques. In the Toffoli decomposition technique we have first identified the total number of multi controlled Toffoli gates required to implement the circuit. Each Toffoli gate is replaced with the equivalent decomposed circuit with intermediate qudit. Total depth and size is of one decomposed circuit is multiplied with the number of Toffoli gates and finally added to the size and depth of the circuit. A n-qubit Toffoli gate can be decomposed into a $\log_2 n$ depth with no ancilla qubits. For example let us assume we need 14 $C^{\otimes 2}$NOT gates, 3 $C^{\otimes 3}$NOT gates and 1 $C^{\otimes 4}$NOT gates. For one $C^{\otimes 2}$NOT gate, both the decomposed circuit size and the depth using intermediate qudit are 3. Therefore both the total size and the depth for implementing 14 $C^{\otimes 2}$NOT gates are 42. Similarly both the total size and the depth for implementing 3 $C^{\otimes 3}$NOT gate are 15. Both the total size and depth for implementing 1 $C^{\otimes 4}$NOT gate are 7. Hence we can conclude that we are getting a exponential improvement over the depth of the circuit. The six tables listed in this section below compare the cost and the depth of the circuit while using standard decomposition and an intermediate qudit decomposition technique. We have taken four different graphs and we find a triangle in every graph. The cells marked with 'yellow' color shows the results of the circuit size and the depth using higher dimensional decomposition technique. \\ Table \ref{W increment oracle} and Table \ref{Dicke incrementoracle} show the circuit parameters for implementing Increment-based oracle using W state and Dicke state, where the number of marked state is chosen as 1 that means we are trying to find one triangle. Hence, while using Dicke or W state the optimal number of iteration will be 1, as $N=4$ and $m=1$. 

\begin{table}[ht!]
\centering
\caption{Total Gate count for Increment-based oracle using W state Preparation}
%\scriptsize
\resizebox{\textwidth}{!}{ 
\begin{tabular}{| c |c | c | c | c | c | c | c |}
 \hline
Decomposition Method used & Type of graph & Number of qubits & 1-qubit gates & 2-qubit gates & Qudit gates & size & Depth  \\ \hline
Standard \cite{sota} & \includegraphics[width=1.5cm, height=1.5cm]{onetriangle.png} & 10 & 314 & 193 & 0 & 507 & 354 \\ \hline
\rowcolor{yellow}Intermediate Qudit & \includegraphics[width=1.5cm, height=1.5cm]{onetriangle.png} & 10 & 33 & 41 & 108 & 182 & 145
\\ \hline
Standard \cite{sota} & \includegraphics[width=1cm, height=1cm]{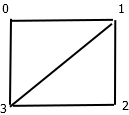} & 10 & 370 & 229 & 0 & 599 & 419\\ \hline
\rowcolor{yellow}Intermediate Qudit & \includegraphics[width=1cm, height=1cm]{squareonediagonal.png} & 10 & 33 & 43 & 126 & 202 & 164\\ \hline
Standard \cite{sota} & \includegraphics[width=1cm, height=1cm]{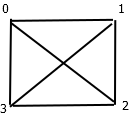} & 10 & 426 & 265 & 0 & 691 & 484 \\
\hline
\rowcolor{yellow}Intermediate Qudit & \includegraphics[width=1cm, height=1cm]{completegraph.png} & 10 & 33 & 45 & 144 & 222 & 183 \\
\hline
Standard \cite{sota} &\includegraphics[width=1cm,height=1cm]{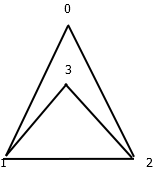} & 10 & 258 & 157 & 0 & 415 & 289 \\
\hline
\rowcolor{yellow}Intermediate Qudit &\includegraphics[width=1cm,height=1cm]{stargraph.png} & 10 & 33 & 44 & 135 & 212 & 174 \\
\hline
\end{tabular}}
\label{W increment oracle}
\end{table}

\hfill
\begin{table}[ht!]
\centering
\caption{Total Gate count for Increment-based oracle using Dicke state Preparation}
%\scriptsize
\resizebox{\textwidth}{!}{ 
\begin{tabular}{| c |c | c | c | c | c | c | c |}
 \hline
Decomposition Method used & Type of graph & Number of qubits & 1-qubit gates & 2-qubit gates & Qudit gates & size & Depth   \\ \hline
Standard \cite{sota} & \includegraphics[width=1.5cm, height=1.5cm]{onetriangle.png} & 10 & 488 & 337 & 0 & 825 & 593 \\ \hline
\rowcolor{yellow}Intermediate Qudit & \includegraphics[width=1.5cm, height=1.5cm]{onetriangle.png} & 10 & 30 & 41 & 108 & 181 & 149
\\ \hline
Standard \cite{sota} & \includegraphics[width=1cm, height=1cm]{squareonediagonal.png} & 10 & 544 & 373 & 0 & 917 & 658\\ \hline
\rowcolor{yellow}Intermediate Qudit & \includegraphics[width=1cm, height=1cm]{squareonediagonal.png} & 10 & 30 & 43 & 126 & 201 & 168\\ \hline
Standard \cite{sota} & \includegraphics[width=1cm, height=1cm]{completegraph.png} & 10 & 600 & 409 & 0 & 1009 & 723 \\
\hline
\rowcolor{yellow}Intermediate Qudit & \includegraphics[width=1cm, height=1cm]{completegraph.png} & 10 & 31 & 45 & 144 & 220 & 189 \\
\hline
Standard \cite{sota} &\includegraphics[width=1cm,height=1cm]{stargraph.png} & 10 & 432 & 301 & 0 & 733 & 528 \\
\hline
\rowcolor{yellow}Intermediate Qudit &\includegraphics[width=1cm,height=1cm]{stargraph.png} & 10 & 30 & 44 & 135 & 209 & 180 \\
\hline
\end{tabular}}
\label{Dicke incrementoracle}
\end{table}
\\
Table \ref{W checking oracle} and Table \ref{Dicke checking oracle} portray the decomposed circuit results while implementing Checking-based oracle using W state and Dicke state respectively.

\begin{table}[h]
\centering
\caption{Total Gate count for Checking-based oracle using W state Preparation}
%\scriptsize
\resizebox{\textwidth}{!}{ 
\begin{tabular}{| c | c | c | c | c | c | c | c |}
 \hline
Decomposition Method used & Type of graph & Number of qubits & 1-qubit gates & 2-qubit gates & Qudit gates & size & Depth  \\ \hline
Standard \cite{sota} & \includegraphics[width=1.5cm, height=1.5cm]{onetriangle.png} & 9 & 324 & 205 & 0 & 529 & 331 \\ \hline
\rowcolor{yellow}Intermediate Qudit & \includegraphics[width=1.5cm, height=1.5cm]{onetriangle.png} & 9 & 34 & 17 & 81 & 131 & 106
\\ \hline
Standard \cite{sota} & \includegraphics[width=1cm, height=1cm]{squareonediagonal.png} & 10 & 396 & 253 & 0 & 649 & 391\\ \hline
\rowcolor{yellow}Intermediate Qudit & \includegraphics[width=1cm, height=1cm]{squareonediagonal.png} & 9 & 33 & 17 & 100 & 150 & 125\\ \hline
Standard \cite{sota} & \includegraphics[width=1cm, height=1cm]{completegraph.png} & 9 & 468 & 301 & 0 & 769 & 471 \\
\hline
\rowcolor{yellow}Intermediate Qudit & \includegraphics[width=1cm, height=1cm]{completegraph.png} & 9 & 34 & 17 & 116 & 166 & 141 \\
\hline
Standard \cite{sota} &\includegraphics[width=1cm,height=1cm]{stargraph.png} & 9 & 252 & 157 & 0 & 409 & 267 \\
\hline
\rowcolor{yellow}Intermediate Qudit &\includegraphics[width=1cm,height=1cm]{stargraph.png} & 9 & 34 & 17 & 100 & 150 & 125 \\
\hline
\end{tabular}}
\label{W checking oracle}
\end{table}

\begin{table}[ht!]
\centering
\caption{Total Gate count for Checking-based oracle using Dicke state Preparation}
%\scriptsize
\resizebox{\textwidth}{!}{ 
\begin{tabular}{| c |c | c | c | c | c | c | c |}
 \hline
Decomposition Method used & Type of graph & Number of qubits & 1-qubit gates & 2-qubit gates & Qudit gates & size & Depth   \\ \hline
Standard \cite{sota} & \includegraphics[width=1.5cm, height=1.5cm]{onetriangle.png} & 10 & 474 & 361 & 0 & 835 & 595 \\ \hline
\rowcolor{yellow}Intermediate Qudit & \includegraphics[width=1.5cm, height=1.5cm]{onetriangle.png} & 10 & 30 & 17 & 107 & 154 & 134
\\ \hline
Standard \cite{sota} & \includegraphics[width=1cm, height=1cm]{squareonediagonal.png} & 10 & 538 & 413 & 0 & 951 & 676\\ \hline
\rowcolor{yellow}Intermediate Qudit & \includegraphics[width=1cm, height=1cm]{squareonediagonal.png} & 10 & 30 & 17 & 100 & 147 & 127\\ \hline
Standard \cite{sota} & \includegraphics[width=1cm, height=1cm]{completegraph.png} & 10 & 600 & 409 & 0 & 1009 & 723 \\
\hline
\rowcolor{yellow}Intermediate Qudit & \includegraphics[width=1cm, height=1cm]{completegraph.png} & 10 & 30 & 17 & 116 & 163 & 143 \\
\hline
Standard \cite{sota} &\includegraphics[width=1cm,height=1cm]{stargraph.png} & 10 & 410 & 309 & 0 & 719 & 518 \\
\hline
\rowcolor{yellow}Intermediate Qudit &\includegraphics[width=1cm,height=1cm]{stargraph.png} & 10 & 30 & 17 & 106 & 153 & 133 \\
\hline
\end{tabular}}
\label{Dicke checking oracle}
\end{table}

Table \ref{Hilbert checking oracle} and Table \ref{Hilbert increment oracle} exhibit the circuit cost while implementing Checking-based oracle and increment-based oracle. The initial state is Full Hilbert space here. So for 4 qubits , $N=16$ and marked state $m=1$. So the optimal number of iteration is 3.

\begin{table}[ht!]
\centering
\caption{Total Gate count for Checking-based oracle in Full Hilbert space}
%\scriptsize
\resizebox{\textwidth}{!}{ 
\begin{tabular}{| c | c | c | c | c | c | c | c |}
 \hline
Decomposition Method used & Type of graph & Number of qubits & 1-qubit gates & 2-qubit gates & Qudit gates & size & Depth  \\ \hline
Standard \cite{sota} & \includegraphics[width=1.5cm, height=1.5cm]{onetriangle.png} & 13 & 1047 & 413 & 0 & 1812 & 331 \\ \hline
\rowcolor{yellow}Intermediate Qudit & \includegraphics[width=1.5cm, height=1.5cm]{onetriangle.png} & 13 & 42 & 48 & 328 & 418 & 381
\\ \hline
Standard \cite{sota} & \includegraphics[width=1cm, height=1cm]{squareonediagonal.png} & 13 & 1239 & 921 & 0 & 2160 & 1416\\ \hline
\rowcolor{yellow}Intermediate Qudit & \includegraphics[width=1cm, height=1cm]{squareonediagonal.png} & 13 & 42 & 48 & 394 & 456 & 419\\ \hline
Standard \cite{sota} & \includegraphics[width=1cm, height=1cm]{completegraph.png} & 13 & 1431 & 1077 & 0 & 2508 & 1659 \\
\hline
\rowcolor{yellow}Intermediate Qudit & \includegraphics[width=1cm, height=1cm]{completegraph.png} & 13 & 42 & 48 & 468 & 558 & 521 \\
\hline
Standard \cite{sota} &\includegraphics[width=1cm,height=1cm]{stargraph.png} & 13 & 855 & 609 & 0 & 1464 & 942 \\
\hline
\rowcolor{yellow}Intermediate Qudit &\includegraphics[width=1cm,height=1cm]{stargraph.png} & 13 & 42 & 48 & 394 & 456 & 419 \\
\hline
\end{tabular}}
\label{Hilbert checking oracle}
\end{table}

\begin{table}[ht!]
\centering
\caption{Total Gate count for Increment-based oracle in Full Hilbert space}
%\scriptsize
\resizebox{\textwidth}{!}{ 
\begin{tabular}{| c | c | c | c | c | c | c | c |}
 \hline
Decomposition Method used & Type of graph & Number of qubits & 1-qubit gates & 2-qubit gates & Qudit gates & size & Depth  \\ \hline
Standard \cite{sota} & \includegraphics[width=1.5cm, height=1.5cm]{onetriangle.png} & 15 & 1731 & 909 & 0 & 2640 & 1284 \\ \hline
\rowcolor{yellow}Intermediate Qudit & \includegraphics[width=1.5cm, height=1.5cm]{onetriangle.png} & 15 & 42 & 175 & 510 & 727 & 688
\\ \hline
Standard \cite{sota} & \includegraphics[width=1cm, height=1cm]{squareonediagonal.png} & 15 & 1599 & 1017 & 0 & 2616 & 1551\\ \hline
\rowcolor{yellow}Intermediate Qudit & \includegraphics[width=1cm, height=1cm]{squareonediagonal.png} & 15 & 42 & 183 & 582 & 807 & 767\\ \hline
Standard \cite{sota} & \includegraphics[width=1cm, height=1cm]{completegraph.png} & 15 & 1767 & 1125 & 0 & 2892 & 1674 \\
\hline
\rowcolor{yellow}Intermediate Qudit & \includegraphics[width=1cm, height=1cm]{completegraph.png} & 15 & 42 & 185 & 600 & 827 & 787 \\
\hline
Standard \cite{sota} &\includegraphics[width=1cm,height=1cm]{stargraph.png} & 15 & 1263 & 801 & 0 & 2064 & 1041 \\
\hline
\rowcolor{yellow}Intermediate Qudit &\includegraphics[width=1cm,height=1cm]{stargraph.png} & 15 & 42 & 183 & 582 & 807 & 767 \\
\hline
\end{tabular}}
\label{Hilbert increment oracle}
\end{table}
As we know the circuit cost is defined by the number of gates used in the circuit and the depth of the circuit is the longest path of the circuit. Performance of the quantum circuit mainly depends on these two parameters. As we can observe from the Table \ref{W increment oracle}, Table \ref{Dicke incrementoracle}, Table \ref{W checking oracle}, Table \ref{Dicke checking oracle}, Table \ref{Hilbert checking oracle} and Table \ref{Hilbert increment oracle}, the oracular circuit is decomposed into 1-qubit and 2-qubit gates as per the state-of-the-art \cite{sota}. The total cost and the depth of are calculated using the conventional decomposition procedure \cite{sota}. Same graph is taken and the circuit is decomposed using intermediate qudits. We observe from the highlighted rows of the tables that the cost and depth of a circuit when it is decomposed in a higher dimensional qudits are sublimer than the conventional approach. For example if we look into the table \ref{Hilbert increment oracle}, for the 1st graph, the size of the graph is 2640 if we use normal decomposition method, but if we apply the intermediate qudit decomposition method, we get the circuit size is 727. So we have a 72\% improvement over the size of the circuit. Similarly the depth of the circuit is 688 using intermediate qudit decomposition as compared to 1284 using conventional decomposition method. So the depth of the circuit is reduced by 46\%. In similar way, we have improved the size and the depth of the circuits for different graphs as shown in the tables.

\subsection{Error Analysis}

Any quantum system is susceptible to different types of errors such as decoherence, noisy gates. For a $d$-dimensional quantum system, the gate error scales as $d^2$ and $d^4$ for 1- and 2-qubit gates respectively \cite{Gokhale_2019}. Furthermore, for qubits, the amplitude damping error decays the state $\ket{1}$ to $\ket{0}$ with probability $\lambda_1$. For a $d$-dimensional system, every state in level $\ket{i} \neq \ket{0}$ has a probability $\lambda_1$ of decaying. In other words, the usage of higher dimensional states penalizes the system with more errors. Nevertheless, the effect of these errors on the used decomposition of Multi-controlled Toffoli gate has been studied by Saha et al. \cite{PhysRevA.105.062453}. They have shown that although the usage of intermediate qudits lead to increased error, the overall error probability of the decomposition is lower than the existing ones since the number of gates and the depth are both reduced. This interpretation is also applicable for our approach of solving clique problem. Hence, we claim that our solution for clique problem with intermediate qudit is more superior in terms of error-efficiency as compared to \cite{sota}. 

\subsection{Time Complexity Analysis}
The time complexity depends on the number of iterations in Grover’s algorithm, the initial state preparation (in case of limited Hilbert space search), different oracles and diffusion operators complexities. The oracle and the diffusion operator are repeated $\frac{\pi}{4}\sqrt{N/M}$  times, which depends on the size of the search space and the expected number of marked states. As detailed in the article \cite{sota} the complexity of the cost of preparing W states is $O(|V|)$ ,the cost of preparing the Dicke state is $O(k|V|)$ for $k$-clique problem. The circuit complexity of W state preparation is $O(log n)$. The complexity of the oracle for the entire Hilbert space is $O(log(k) + |E| + |V|)$. The complexity of the oracle when using initial state preparation to limit the search space is $O(log(k) + |E|)$. The total complexity is $O(\sqrt{\frac{{n \choose k}}{m}} \times (O(oracle) + O(diffusion operator)))$, where $m =$ marked states and $k =$ clique size. If we use W state or Dicke state as initial state preparation, the complexity becomes
$O(State preparation)+ O(\sqrt{\frac{{n \choose k}}{m}} \times (O(oracle) + O(diffusion operator)))$. The run-time complexity of the algorithm is reduced to logarithmic factor due to the Toffoli decomposition method applied to the circuit. As Maximum Clique Problem is an iterative problem and it depends on the size of $k$ we have not showed the time complexity of this problem.

\section{Conclusion}
In this article, we have proposed an improved quantum circuit for the clique problem with the help of higher-dimensional intermediate temporary qudits. We have analyzed the performance of the proposed approaches and the state-of-the-art approach from different perspectives, such as gate count, circuit size and circuit depth. Effect of error has also been analyzed on the circuit for using higher dimensional gates. The result of the cost analysis and error analysis shows that the circuit can have a lower gate count and depth than the state-of-the-art approach which leads to the lower error probability. The optimized circuit can be implemented in real quantum devices with lower depth and cost. The results are very promising which pave the way for circuit implementation of other computational problems in near future.



%% References


\bibliographystyle{elsarticle-num}
\bibliography{biblio}






% \end{thebibliography}

\end{document}